\documentclass[conference]{IEEEtran}
\IEEEoverridecommandlockouts
\usepackage{caption}
\usepackage{cite}
\usepackage{hyperref}
\usepackage{amsmath,amssymb,amsfonts}
\usepackage{algorithmic}
\usepackage{graphicx}
\usepackage{textcomp}
\usepackage{xcolor}
\usepackage{float}
\usepackage{placeins}
\usepackage{amsmath}

\def\BibTeX{{\rm B\kern-.05em{\sc i\kern-.025em b}\kern-.08em
    T\kern-.1667em\lower.7ex\hbox{E}\kern-.125emX}}

\usepackage{titlesec}
\titlespacing*{\section}{0pt}{*2}{*1}
\titlespacing*{\subsection}{0pt}{*1.5}{*0.5}

\begin{document}

\title{AutoPipelineAI: Context-Aware CI/CD Pipeline Generation from Natural Language \\
}
\author{
  \IEEEauthorblockN{Youssef Mohamed Aboelfotoh}
  \IEEEauthorblockA{Faculty of Computing and AI \\
    Cairo University \\
    Cairo, Egypt \\
    youssefmohamed81222@gmail.com}
  \and
  \IEEEauthorblockN{Mohamed Ahmed Hemdan}
  \IEEEauthorblockA{Faculty of Computing and AI \\
    Cairo University \\
    Cairo, Egypt \\
    hemdan.swe@gmail.com}
  \and
  \IEEEauthorblockN{Mohammad El-Ramly}
  \IEEEauthorblockA{Faculty of Computing and AI \\
    Cairo University \& University of \\ 
    London Branch (EUE), Egypt \\
    m.elramly@fci-cu.edu.eg\\ORCID 0000-0002-5076-3829}
  
  \and
  \IEEEauthorblockN{Khlood Hassan}
  \IEEEauthorblockA{Faculty of Computing and AI \\
    Cairo University \\
    Cairo, Egypt \\
    khlood.hassan74@gmail.com}
\and
  \IEEEauthorblockN{Mahmoud Saleh Saad}
  \IEEEauthorblockA{Faculty of Computing and AI \\
    Cairo University \\
    Cairo, Egypt \\
    mahmoud22saleh22@gmail.com}
  \and
  \IEEEauthorblockN{Ahmed Mohamed Tolba}
  \IEEEauthorblockA{Faculty of Computing and AI \\
    Cairo University \\
    Cairo, Egypt \\
    bassuonybassuony6@gmail.com}

  \and
  \IEEEauthorblockN{Seif Gamal Abdelmonem}
  \IEEEauthorblockA{Faculty of Computing and AI \\
    Cairo University \\
    Cairo, Egypt \\
    seifgamalrayan@gmail.com}
}

\IEEEpubid{\makebox[\columnwidth]{  \hfill}
\hspace{\columnsep}\makebox[\columnwidth]{}}
\maketitle
\IEEEpubidadjcol

\begin{abstract}
Modern software development relies on CI/CD pipelines to automate testing, building, and deployment operations. Configuring DevOps pipelines is challenging and time-consuming, as developers must understand platform-specific syntax and manually create configuration files. This complexity can lead to configuration errors and reduced productivity, especially for developers with limited DevOps experience. This paper introduces the AutoPipelineAI system, which generates CI/CD pipeline configurations using natural language descriptions. The proposed solution uses large language models (LLMs) to translate developer intent, analyze repository structures, and create specific pipeline scripts for environments like GitHub Actions and GitLab CI/CD. It integrates repository-aware analysis, automated validation systems, and a feedback mechanism that confirms the accuracy and usability of the created pipelines. We present the system architecture, its implementation, and an assessment framework designed to measure generation precision, configuration validity, and reduction in setup effort compared to manual pipeline creation. AutoPipelineAI illustrates how LLMs can simplify the complexity of DevOps configuration and enhance developer access to continuous delivery methods. Evaluation results provide early evidence that repository-aware, natural-language-driven CI/CD generation is a viable and promising paradigm for reducing the complexity of DevOps configuration and enabling more accessible software delivery automation.
\end{abstract}

\begin{IEEEkeywords}
DevOps, CI/CD, Large Language Models, Natural Language Processing, Software Automation, Pipeline Generation, GitHub Actions, Software Engineering
\end{IEEEkeywords}
\section{Introduction}
Continuous Integration / Continuous Deployment (CI/CD) pipelines are essential elements of contemporary software engineering, enabling developers to enhance the efficiency of testing, building, and deployment workflows. By automating repetitive tasks, these pipelines improve software reliability and accelerate delivery cycles while minimizing the risk of human error. However, despite these benefits, creating and maintaining these configurations remains a difficult and time-consuming process. Developers often need to master platform-specific syntax, manage configuration files, and integrate third-party tools, creating a steep learning curve that frequently leads to mistakes, especially for entry-level developers or teams unfamiliar with DevOps practices.

Current solutions address these problems partially through AI-powered recommendations, debugging assistants, or template-based configurations within CI/CD systems. For instance, platforms like CircleCI offer natural-language support for troubleshooting, while observability tools utilize AI to detect anomalies in deployment logs. Despite these advancements, a significant gap remains: existing tools do not support the end-to-end generation of comprehensive, repository-aware pipelines directly from high-level natural language descriptions.

To fill this gap, we propose AutoPipelineAI, a context-aware framework for generating CI/CD pipelines from natural language using large language models. The motivation for developing this system is to democratize infrastructure automation by removing the syntax barrier, allowing developers to focus on core logic while the AI handles the complex "plumbing" of the pipeline. AutoPipelineAI achieves this by integrating repository-aware analysis with Large Language Models (LLMs) to translate developer intent into validated, ready-to-use pipeline scripts tailored to specific environments.

The rest of this paper is organized as follows: Section 2 provides the necessary background on CI/CD and LLMs; Section 3 reviews related work in automated DevOps; Section 4 is the problem statement addressed by our work; Section 5 details our research methodology, including the system architecture and implementation; Section 6 presents the evaluation of the system using our assessment framework; and Section 7 provides the conclusion and suggestions for future work.

\section{Background}\label{sec2}
 This section introduces the key concepts and ideas that form the basis of the AutoPipelineAI’s design and implementation.
\subsection{ \textbf{Continuous Integration and Continuous Deployment}}\label{subsec1}
Continuous Integration (CI) is a software development practice that involves engineers regularly integrating code changes into a shared repository, typically multiple times a day. To detect errors early in the development cycle, each integration triggers an automated build and testing procedure. Continuous Deployment (CD) extends this practice by automatically deploying any code changes that pass the automated testing phase directly to production environments.
\subsection{\textbf{CI/CD Platform Configuration}}\label{subsec2}
Modern CI/CD platforms require developers to define workflows using platform-specific syntax. GitHub Actions~\cite{b8} utilizes YAML files stored in \texttt{.github/workflows/}, GitLab CI/CD~\cite{b9} employs \texttt{.gitlab-ci.yml}, and Jenkins~\cite{b11} relies on Groovy-based \textit{Jenkinsfiles}. Each platform maintains distinct syntax conventions, environment variable handling mechanisms, and integration patterns with external services. Creating these configurations requires platform expertise and careful adherence to syntax and best practice

\subsection{\textbf{Large Language Models in Software Engineering}}\label{subsec3}
Large Language Models (LLMs) such as GPT-4~\cite{b7}, Claude, and Gemini have demonstrated remarkable capabilities in understanding and generating code across multiple programming languages. These models are trained on vast corpora of text and code, enabling them to perform tasks such as code completion, documentation generation, bug detection, and natural language to code translation. In the context of DevOps automation, LLMs can interpret developer intent expressed in natural language and translate it into structured configuration files. By leveraging their understanding of both natural language semantics and programming syntax, LLMs can bridge the gap between high-level workflow descriptions and low-level implementation details.

\subsection{\textbf{ Repository Analysis and Context Awareness}}\label{subsec4}
Effective pipeline generation requires understanding structure and characteristics of the target repository. Key indicators are:
\begin{itemize}
    \item \textbf {Package managers:} Files like package.json (Node.js), requirements.txt (Python), pom.xml
(Java), or Cargo.toml (Rust) indicate the project’s
language and dependency management approach.
\end{itemize}
\begin{itemize}
    \item \textbf{Build tools:} Presence of Makefile, build.gradle, or CMakeLists.txt suggests specific build processes
\end{itemize}
\begin{itemize}
    \item \textbf{Containerization:} Dockerfile or docker-compose.yml files indicate Docker-based workflows.
\end{itemize}
\begin{itemize}
    \item \textbf{Testing frameworks:} Test directories and configuration files reveal the project’s testing strategy
\end{itemize}
\begin{itemize}
  \item \textbf{Environment configuration:} env.example or configuration files indicate required environment variables. \end{itemize}
  By analyzing these artifacts, an intelligent pipeline generator can produce context-appropriate configurations aligned with the project’s toolchain and development practices.

\subsection{\textbf{YAML Validation and Schema Compliance}}\label{subsec5}
CI/CD configuration files must adhere to strict schema
requirements defined by each platform to ensure executable, reliable configurations. Invalid syntax, incorrect structure, or semantic analysis leads to pipeline failures.

\section{Related Work}\label{sec3}
The automation of CI/CD pipeline configuration has been
explored through several approaches, ranging from traditional
template-based systems to AI-assisted tools. This section
categorizes existing work and highlights the limitations that
AutoPipelineAI addresses.
\subsection{\textbf{Traditional and AI-Assisted DevOps Approaches}}\label{subsec1}
CI/CD plays an important role in modern software development because it improves automation and delivery speed, especially in large and multi-team projects, as highlighted by Kisina et al. ~\cite{b1} However, platforms like GitHub Actions ~\cite{b8}, GitLab CI/CD ~\cite{b9}, Jenkins ~\cite{b11}, and CircleCI provide powerful automation features, but pipeline configuration still often requires manual effort. To address this, Labonté-Lamoureux and Boyer ~\cite{b15} found that using templates and self-service portals brings consistency and saves time. While Donca et al. ~\cite{b16} demonstrated that automated pipeline generation can accelerate deployment and enhance standardization. They achieved pipeline execution in 157 seconds and decreased infrastructure costs by 80-85\% compared to doing it manually. More recently, AI-enhanced DevOps automation, monitoring, and decision-making was demonstrated by Vemuri et al  \cite{b2} and Oyeniran et al ~\cite{b3}. Tools such as CircleCI and Gemini Code Assist use AI to improve debugging and productivity in CI/CD operations. However, rather than creating new pipelines, these methods optimize current ones. As a result, there is still a gap in designing intelligent systems that can generate CI/CD pipelines directly from natural language inputs.
\subsection{\textbf { LLM Evaluation and Testing Frameworks}}\label{subsec2}
Recent studies in 2025 by Baqar et al~\cite{b4}, Patel et al~\cite{b5}, and Cui et al~\cite{b6} highlight the growing use of generative AI and LLMs to enhance DevOps automation, decision-making, and pipeline optimization. In addition, tools such as Promptfoo ~\cite{b12} and DeepEval~\cite{b13} support the evaluation of AI models within CI/CD workflows. However, these approaches focus mainly on improving and monitoring existing pipelines rather than creating them. As a result, they still depend on manually pre-configured workflows, revealing a gap in automatic CI/CD pipeline generation from high-level or natural language inputs.
\subsection{\textbf { Research Gaps and Positioning}}\label{subsec3}
Despite these advancements, no existing solution integrates natural language interpretation, repository-aware analysis, multi-platform generation, and automated validation into a cohesive pipeline creation framework. Current tools generally remain restricted to manual configuration, post-execution debugging assistance, or testing-focused evaluation frameworks. AutoPipelineAI is a first attempt to bridge these gaps by synthesizing functional CI/CD pipelines directly from high-level natural language descriptions. Unlike static template-driven approaches, AutoPipelineAI performs structural analysis of the repository to ensure configurations are contextually accurate. Furthermore, while most AI-based DevOps tools are reactive, AutoPipelineAI is proactive, generating end-to-end workflows from the outset. By unifying LLM-driven intent interpretation \cite{b7, b8, b10}, deep repository scanning, and automated validation, AutoPipelineAI is a step in DevOps automation that reduces cognitive load and democratizes access to continuous delivery for developers of all experience levels.
\section{Problem Statement}\label{sec4}
Despite the widespread adoption of CI/CD, the manual configuration of these pipelines introduces significant developmental overhead. Developers frequently spend more time managing platform-specific YAML or JSON files than focusing on core application logic. These configurations are notoriously fragile; even minor syntax or indentation errors can lead to total pipeline failure, creating a bottleneck in the delivery lifecycle.

Furthermore, the CI/CD landscape is highly fragmented, with each platform (e.g., GitHub Actions, GitLab CI/CD, Jenkins) requiring mastery of distinct syntax, trigger logic, and environment conventions. This fragmentation creates a steep learning curve, even for experienced engineers, and often results in sub-optimal, "copy-pasted" configurations that lack project-specific optimization. Currently, there is a lack of automated solutions capable of bridging the gap between high-level human intent and low-level infrastructure code. There is a critical need for an intelligent interface that leverages natural language to abstract the complexities of DevOps syntax, allowing developers to generate valid, repository-aware configurations through intuitive descriptions.

\section{Research Methodology}\label{sec5}

The architectural design of AutoPipelineAI is built upon a modular, multi-layered framework designed to facilitate the automated transition from natural language intent to functional infrastructure code. As illustrated in Fig. 1, the system is partitioned into distinct tiers that prioritize operational resilience and the decoupling of complex computational logic from user interactions. The following subsections detail the functional hierarchy of these layers and the orchestration strategies employed to ensure consistent pipeline generation.

\captionsetup{margin={1em,0em}}

\begin{figure}[htbp]
\centering
\includegraphics[width=\columnwidth,height=10cm]{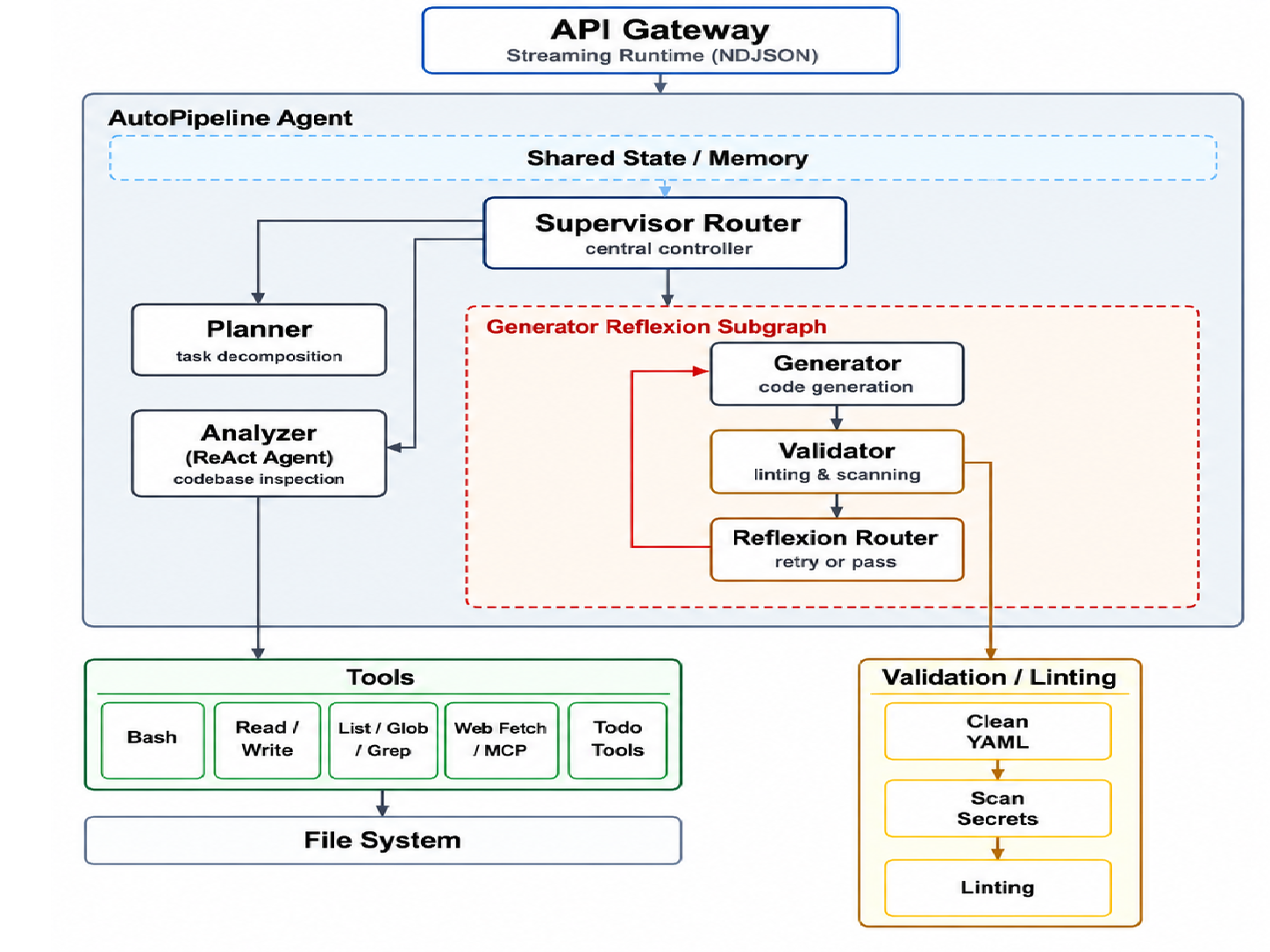}
\caption{Architecture of the AutoPipelineAI system,
showing the interconnected components for workflow generation,
processing, and deployment.}
\label{fig1}
\end{figure}

\subsection{\textbf{Agent Processing Layer}}
The AutoPipelineAI system incorporates a dedicated multi-agent processing layer responsible for repository-aware CI/CD workflow generation. The agent architecture follows a graph-oriented orchestration model in which specialized agents collaborate through shared state and controlled execution routing. As illustrated in Fig. 1, the agent subsystem is organized as a coordinated multi-agent workflow in which specialized agents collaborate through shared state propagation and iterative validation-driven execution. We further explain these agents.

\begin{itemize}
    \item \textbf{Supervisor Router:}
    The supervisor router acts as the central coordination component of the agent pipeline. It manages execution flow between agents, maintains shared context, and determines task delegation strategies based on user requests and repository state.

    \item \textbf{Shared State and Memory:}
    A shared state layer enables information exchange between agents during execution. Repository metadata, intermediate workflow representations, validation outputs, and refinement history are propagated through this shared memory structure to ensure consistency across iterative processing stages.

    \item \textbf{Planner Agent:}
    The planner agent performs task decomposition by transforming high-level natural language requirements into structured workflow objectives. This stage identifies required pipeline stages, deployment targets, testing requirements, and platform-specific constraints.

    \item \textbf{Repository Analyzer Agent:}
    The analyzer component follows a repository-aware ReAct-style strategy to inspect project structure and infer development context. The agent examines dependency manifests, build scripts, containerization files, and testing configurations to identify the technologies and operational requirements associated with the target repository.

    \item \textbf{Generator–Validator Reflexion Subgraph:}
    The workflow synthesis process is implemented as an iterative reflexion subgraph composed of three tightly coupled components:
    \begin{itemize}
        \item \textbf{Generator Agent:}
        The generator agent produces CI/CD configuration files based on the inferred repository context and developer intent. Generated workflows include platform-specific syntax, dependency management steps, build stages, testing procedures, and deployment instructions.
        \item \textbf{Validator Agent:}
        The validator agent performs structural and semantic validation of generated configurations. Validation includes YAML formatting checks, schema verification, linting operations, and secret scanning to detect configuration vulnerabilities or invalid workflow definitions.
        \item \textbf{Reflexion Router:}
        The reflexion router evaluates validator outputs and determines whether the generated workflow satisfies correctness constraints. Invalid or incomplete configurations are routed back to the generator for iterative refinement until validation requirements are satisfied.
    \end{itemize}
    This reflexion-based execution model improves generation robustness by introducing automated self-correction loops into the pipeline synthesis process.

    \item \textbf{Tool Integration Layer:}
    The agent subsystem is connected to a tool execution layer that provides controlled access to filesystem operations, shell execution, repository inspection, web retrieval, and auxiliary utilities. Tool augmentation enables the agents to ground reasoning in repository-specific contextual information rather than relying solely on prompt-based inference.
\end{itemize}

Ultimately, this modular architecture directly addresses the core challenges of CI/CD configuration by abstracting DevOps complexity behind a natural-language-driven framework. The multi-agent design translates developer intent into platform-specific configurations without requiring mastery of underlying YAML syntax. Crucially, the iterative Generator–Validator Reflexion loop mitigates the fragility of manual configuration by proactively detecting and correcting schema and logic errors prior to deployment. By unifying structural repository analysis, automated validation, and resilient agent orchestration, AutoPipelineAI bridges the gap between high-level requirements and executable infrastructure, significantly reducing development overhead.

\section{Evaluation}\label{sec6}

The objective of this evaluation is to assess the capability of AutoPipelineAI, using LLMs, to generate valid and functionally equivalent CI/CD workflows across multiple projects.

\subsection{\textbf{Experimental Design}}\label{sec1}

For evaluation, we used two  systems as our dataset \cite{b14}.
\begin{itemize}
\item Project 1: Peewee is a medium-scale Python project featuring a moderately complex CI/CD workflow.
\item Project 2: Hugo is a large-scale Go project with a highly complex CI/CD workflow.
\end{itemize}

These projects were selected to reflect different levels of CI/CD complexity and to provide diverse evaluation contexts. The ground truth was considered the YAML files coming with the project and our system generated its own CI/D pipeline. Pipeline generation was then tested using four  LLMs:
\begin{itemize}
    \item GPT-4.1
    \item GPT-4o
    \item MiniMax-M2-1
    \item Qwen3-Coder-Next-UD-79B
\end{itemize}

The evaluation criteria were organized into two complementary phases designed to measure both the semantic quality and the functional correctness of the generated workflows.
\subsubsection{\textbf{AI-assisted evaluation framework}} 
The first phase is to use ChatGPT as evaluator in an AI-assisted evaluation framework to evaluate the generated YAML outputs based on predefined evaluation criteria, including accuracy and completeness. To ensure standardization during evaluation, a uniform prompting technique was applied across the models for consistency and fairness. This stage focuses on how well the generated workflow aligned with the expected implementation framework and needed project functionality. Table~\ref{tab:multi_project} presents the evaluation results of four models across two distinct projects developed in Python and Go. 

Let $Y^{*}_j$ denote the ground-truth YAML for project $j$, and $\hat{Y}_{i,j}$ denote the workflow generated by model $i$ for project $j$, where $j \in \{1,2\}$. Let $C(Y)$ denote the set of workflow components and $S(Y)$ the set of required steps.

\paragraph{Accuracy}
\begin{equation}
\text{Accuracy}_{i,j} = \frac{|C(Y^{*}_j) \cap C(\hat{Y}_{i,j})|}{|C(Y^{*}_j)|}
\end{equation}

\paragraph{Completeness}
\begin{equation}
\text{Completeness}_{i,j} = \frac{|S(\hat{Y}_{i,j})|}{|S(Y^{*}_j)|}
\end{equation}

\paragraph{Aggregate Score (overall performance across projects)}
\begin{equation}
\bar{M}_i = \frac{1}{2} \sum_{j=1}^{2} M_{i,j}
\end{equation}

\begin{table*}[htbp]
\caption{Multi-project evaluation of generated CI/CD workflows (Python and Go).}
\centering
\small
\begin{tabular}{lccccc}
\hline
Model 
& \multicolumn{2}{c}{Python Project} 
& \multicolumn{2}{c}{Go Project} 
& Mean \\
\cline{2-5}
& Acc & Comp  
& Acc & Comp 
& Acc \\
\hline

GPT-4.1 
& 0.76 & 0.82 
& 0.89 & 0.82 
& 0.83 \\

GPT-4o 
& 0.76 & 0.78  
& 0.78 & 0.75  
& 0.77 \\

MiniMax-M2-1 
& 0.84 & 0.88  
& 0.86 & 0.92 
& 0.85 \\

Qwen3-Coder-Next-UD-79B 
& 0.84 & 0.88  
& 0.90 & 0.95 
& 0.87 \\

\hline
\end{tabular}
\vspace{2mm} 

\label{tab:multi_project}

\end{table*}

\subsubsection{\textbf{Requirement-based Functional Evaluation}} 
During the second phase, a functional assessment based on requirements was performed using repository-specific validation. Every generated workflow was manually and programmatically examined against a predefined checklist of functional requirements derived from target repository specifications. Based on this process, the functional evaluation summary for Peewee is presented in Table ~\ref{tab:peewee_functional_eval}, while results for Hugo are shown in Table ~\ref{tab:hugo_functional_eval}. Likewise, the error type distribution for Peewee is reported in Table ~\ref{tab:peewee_error_distribution}, whereas results for Hugo are provided in Table ~\ref{tab:hugo_error_distribution}. Finally, Table ~\ref{tab:overall_model_performance} presents overall model performance across both projects.

\begin{table*}[htbp]
\centering
\caption{Peewee (Python) Functional Evaluation Summary}
\label{tab:peewee_functional_eval}
\renewcommand{\arraystretch}{1.2}
\begin{tabular}{|p{4.5cm}|c|c|c|c|}
\hline
\textbf{Requirement} & \textbf{GPT-4.1} & \textbf{GPT-4o} & \textbf{MiniMax} & \textbf{Qwen3} \\
\hline
MariaDB DB Name & Fail & Fail & Fail & Fail \\
\hline
MariaDB Health Check & Fail & Partial & Fail & Fail \\
\hline
Python Dependencies & Partial & Full Pass & Fail & Fail \\
\hline
Test Execution (--engine flag) & Fail & Full Pass & Fail & Fail \\
\hline
Test CLI Flags (psql/mysql) & Fail & Fail & Fail & Fail \\
\hline
CockroachDB Path & Full Pass & Full Pass & Fail & Full Pass \\
\hline
Branch Trigger Wildcard & Fail & Fail & Fail & Fail \\
\hline
Raw Pass Rate & 21\% & 50\% & 0\% & 14\% \\
\hline
Fixes Required & 6/7 & 4/7 & 7/7 & 6/7 \\
\hline
\end{tabular}

\vspace{0.3cm}
\small
\textbf{Legend:} Full Pass = 1.0, Partial = 0.5, Fail = 0
\end{table*}

\begin{table*}[htbp]
\centering
\caption{Peewee (Python) Error Type Distribution}
\label{tab:peewee_error_distribution}
\renewcommand{\arraystretch}{1.2}
\begin{tabular}{|p{5cm}|c|c|c|c|}
\hline
\textbf{Error Type} & \textbf{GPT-4.1} & \textbf{GPT-4o} & \textbf{MiniMax} & \textbf{Qwen3} \\
\hline
Hallucination & 0 & 0 & 1 & 2 \\
\hline
Configuration Error & 2 & 1 & 2 & 2 \\
\hline
Silent Logic Error & 1 & 0 & 1 & 1 \\
\hline
Omission & 2 & 2 & 1 & 1 \\
\hline
Syntax Error & 1 & 1 & 1 & 1 \\
\hline
Environment Unawareness & 0 & 0 & 1 & 1 \\
\hline
\textbf{Total Errors} & \textbf{6} & \textbf{4} & \textbf{7} & \textbf{8} \\
\hline
\end{tabular}
\end{table*}

\begin{table*}[htbp]
\centering
\caption{Hugo (Go) Functional Evaluation Summary}
\label{tab:hugo_functional_eval}
\renewcommand{\arraystretch}{1.2}
\begin{tabular}{|p{5cm}|c|c|c|c|}
\hline
\textbf{Requirement} & \textbf{GPT-4.1} & \textbf{GPT-4o} & \textbf{MiniMax} & \textbf{Qwen3} \\
\hline
Job Timeout & Fail & Fail & Fail & Full Pass \\
\hline
Disk Space Management & Fail & Fail & Fail & Fail \\
\hline
Dart Sass Checksums & Fail & Fail & Full Pass & Fail \\
\hline
Sass Hash Verification Logic & Fail & Fail & Fail & Fail \\
\hline
OS Shell Isolation & Full Pass & Fail & Fail & Full Pass \\
\hline
Goat CLI Repository & Fail & Fail & Fail & Full Pass \\
\hline
Node.js Version (v22) & Fail & Fail & Fail & Fail \\
\hline
TestNPMGlobalInstalls Skip & Fail & Fail & Fail & Fail \\
\hline
DragonflyBSD Build Command & Fail & Fail & Full Pass & Partial \\
\hline
Staticcheck Cache Management & Fail & Fail & Fail & Fail \\
\hline
Raw Pass Rate & 10\% & 0\% & 20\% & 35\% \\
\hline
Fixes Required & 9/10 & 10/10 & 8/10 & 7/10 \\
\hline
\end{tabular}

\vspace{0.3cm}
\small
\textbf{Legend:} Full Pass = 1.0 \; | \; Partial = 0.5 \; | \; Fail = 0
\end{table*}

\begin{table*}[htbp]
\centering
\caption{Hugo (Go) Error Type Distribution}
\label{tab:hugo_error_distribution}
\renewcommand{\arraystretch}{1.2}
\begin{tabular}{|p{5cm}|c|c|c|c|}
\hline
\textbf{Error Type} & \textbf{GPT-4.1} & \textbf{GPT-4o} & \textbf{MiniMax} & \textbf{Qwen3} \\
\hline
Hallucination & 1 & 2 & 2 & 2 \\
\hline
Configuration Error & 2 & 2 & 1 & 1 \\
\hline
Resource Management & 2 & 2 & 2 & 2 \\
\hline
OS Compatibility & 0 & 2 & 1 & 0 \\
\hline
Omission & 2 & 1 & 1 & 2 \\
\hline
\textbf{Total Errors} & \textbf{7} & \textbf{9} & \textbf{7} & \textbf{7} \\
\hline
\end{tabular}
\end{table*}

\begin{table*}[htbp]
\centering
\caption{Overall Functional Evaluation Across Both Projects}
\label{tab:overall_model_performance}
\renewcommand{\arraystretch}{1.2}
\begin{tabular}{|p{2cm}|c|c|c|c|c|}
\hline
\textbf{Model} & \textbf{Peewee Raw Pass Rate} & \textbf{Hugo Raw Pass Rate} & \textbf{Micro-Average} & \textbf{Total Errors} & \textbf{Hallucinations} \\
\hline
GPT-4.1 & 21\% & 10\% & 15\% & 13 & 1 \\
\hline
GPT-4o & 50\% & 0\% & 21\% & 13 & 2 \\
\hline
MiniMax & 0\% & 20\% & 12\% & 14 & 3 \\
\hline
Qwen3 & 14\% & 35\% & 26\% & 15 & 4 \\
\hline
\end{tabular}
\end{table*}

\subsection{\textbf{Discussion}}

This research employed two complementary evaluation perspectives: a functional requirement–based assessment and an AI-based quality evaluation. The results indicate that these approaches capture different dimensions of CI/CD workflow generation and should therefore be interpreted jointly rather than independently. The AI-based evaluation further revealed that model performance varies across projects with different levels of complexity. While most models achieved reasonable structural accuracy and generated workflows that appeared well organized, they often struggled with dependency management and execution reliability. These findings suggest that structural similarity to reference workflows does not necessarily translate into operational correctness or successful execution.

The Hugo complex project exposed notable weakness in consistency, dependency management, and timeout configuration. In contrast, simpler workflows generally received strong scores in both structural quality and functional correctness. These findings indicate that workflow complexity remains a major obstacle for existing LLM-driven DevOps automation. Among evaluated models, Qwen3-Coder-Next-UD-79B exhibited relatively consistent performance across both repositories, indicating greater ability to generalize. MiniMax-M2-1 also showed competitive performance in structural completeness and organization. In contrast, GPT-4o tended to be less consistent, particularly when handling complex workflows.

The functional evaluation reveals three consistent patterns across all evaluated models and repositories.
\textbf{Silent logic errors:} represent the most common failure category. In the Peewee workflow, omission of –engine flag in three of the four models caused all matrix jobs to execute against SQLite regardless of the configured backend, producing a green pipeline with entirely incorrect results. Similar behavior appeared in the Hugo project where incorrect timeout values caused silent termination. These failures cannot be detected by pipeline status monitoring alone.
\textbf{Hallucination of external references:} appeared in three of four models. MiniMax generated a non-existent PyPI package (psycopg3-binary), Qwen3 produced placeholder SHA256 checksums, and GPT-4.1, GPT-4o, and MiniMax all attributed the Goat CLI tool to incorrect GitHub repositories. These errors are syntactically feasible, making them undetectable without execution or expert review.
\textbf{Universal configuration failures:} appeared regardless of model architecture. In the Peewee repository, every model evaluated produced errors related to MariaDB database naming or branch-trigger wildcard configuration. Every model in the Hugo project produced insufficient disk cleanup and inadequate timeout values. These patterns suggest systematic gaps in repository-specific and platform-specific knowledge.
In general, current LLMs are stronger in creating structurally sound CI/CD workflows rather than reliable production-ready pipelines. Performance also varied by repository: GPT-4o performed better on Peewee but weakest on Hugo, while Qwen3-Coder-Next-UD-79B demonstrated improved results on Hugo. This suggests domain-dependent capability rather than universally reliable CI/CD reasoning

Taken together, these findings highlight both the promise and limitations of current LLMs in CI/CD pipeline generation. While models consistently produce syntactically valid workflows, they remain prone to silent logic errors, hallucinated references, and universal configuration gaps that undermine reliability. The variation in performance across repositories further underscores that capability is domain‑dependent rather than universally robust. These results suggest that advancing toward production‑ready automation will require integrating LLMs with stricter schema validation, repository‑aware knowledge, and expert oversight to bridge the gap between structural soundness and operational dependability.

\subsection{\textbf{Limitations}}

This study has several limitations. 
 \textbf{First}, only two repositories and four models were evaluated. The chosen projects are different in language and complexity, larger experiments are necessary for generalization.
\textbf{Second}, the functional evaluation relied on repository-specific requirements and some manual judgment, which might lead to subjective evaluation even with established scoring criteria.
\textbf{Third}, AI driven evaluation relies on the reasoning patterns of the evaluator model. Such evaluation may be influenced by prompting wording or evaluator bias
\textbf{Finally}, the two evaluation methods evaluate distinct constructs-functional accuracy versus perceived quality, and their results should not be seen as directly interchangeable.

\section{Conclusion}
This study introduced AutoPipelineAI, a context-aware system for generating CI/CD pipelines from natural language using large language models (LLMs). We evaluated four state-of-the-art models—GPT-4.1, GPT-4o, MiniMax-M2-1, and Qwen3-Coder-Next-UD-79B—across workflows of different complexity, including a Python-based system (Peewee) and a Go-based static site pipeline (Hugo).

Overall, the results provide encouraging evidence that LLMs can serve as a practical first step toward generating CI/CD pipelines directly from natural language descriptions. This supports the vision of reducing the DevOps expertise required to create and maintain automation workflows, making CI/CD more accessible to a broader range of developers. However, the findings also show that while LLMs are generally effective at producing the overall structure of CI/CD workflows, their reliability decreases as workflow complexity increases. In other words, the more multi-stage and interconnected a pipeline becomes, the more likely models are to introduce errors in logic, dependencies, or execution behavior. Furthermore, performance varies considerably across models, highlighting the importance of model selection when deploying LLM-assisted DevOps solutions.

Across both systems used for evaluation, we consistently observed three main types of issues. First, silent logic errors—cases where pipelines look correct and even pass checks, but actually do the wrong thing during execution. These are especially dangerous because they are hard to notice from CI status alone. Second, hallucinated dependencies and external references, where models generate packages, repositories, or checksums that look real but do not exist. Third, we found that workflow complexity strongly affects performance, with all models struggling more as pipelines became larger and more multi-stage.

We also noticed a clear “complexity gap.” Models that performed reasonably well on simpler workflows often struggled with more advanced configurations. Among the tested models, MiniMax-M2-1 was relatively consistent in producing well-structured outputs, while GPT-4o became less stable as complexity increased. Qwen3-Coder-Next-UD-79B stood out for maintaining consistent performance across both projects, suggesting better adaptability to different DevOps setups.

In summary, the key takeaway is that CI/CD pipelines generated by LLMs can look correct on the surface but still fail in important ways when executed. This gap between “looking right” and “working” becomes more noticeable as workflow complexity increases, making complexity a key factor in how reliable these models are in real DevOps environments.
\subsection{\textbf{Implications}}
These findings indicate that LLM-driven CI/CD pipeline creation is a promising approach. But it should not be considered completely self-sufficient. Instead, generated pipelines should be considered draft artifacts requiring validation, execution testing, and domain-specific review.
\subsection{\textbf{Future Work}}
This research shows that workflow complexity is the main factor influencing how well LLMs can generate infrastructure-as-code. To close the remaining reliability gap, future work will focus on four directions:
\begin{itemize}
    \item \textbf{Hybrid Synthesis:} Combining the creative reasoning of LLMs with rule-based validation engines to guarantee syntax correctness.
    \item \textbf{Runtime-Aware Feedback:} Using live execution logs from platforms such as \textit{GitHub Actions} so agents can automatically detect and fix failed deployments.
    \item \textbf{Cross-Domain Generalization:} Extending evaluation to polyglot repositories and multi-cloud targets to test adaptability in diverse environments.
    \item \textbf{Retrieval-Augmented Generation (RAG):} Grounding pipeline generation in real repository artifacts and external documentation to reduce hallucinations and improve reliability.
\end{itemize}

By addressing these challenges, we aim to transition CI/CD automation from high-level "drafting" to fully autonomous, production-grade infrastructure generation.

\section*{Acknowledgment}
The authors express gratitude to Siemens Egypt for their technical mentorship and guidance throughout the project.

\end{document}